%% file: paper.tex
\documentclass[letterpaper, 10 pt, conference]{llncs}

\usepackage{microtype}
\usepackage{amsmath, amsfonts}
\usepackage[utf8]{inputenc}
\usepackage{todonotes}
\usepackage{tikz}
\usepackage{cite}
\usepackage{multirow}
\usepackage{hyperref}
\usepackage{listings}
\usepackage[noeepic]{qtree}
\usepackage[caption=false]{subfig}
\usepackage{tabularx}
\usepackage{xspace}
\usepackage{wrapfig}
\usepackage{microtype}
\usepackage[linesnumbered,ruled,vlined]{algorithm2e}

\SetCommentSty{mycommfont}

\usetikzlibrary{shapes,arrows, calc, automata, fit}

\newcommand{\toolname}{\textsc{SynRG}\xspace}

\input{commands.tex}
\lstdefinelanguage{alg}{
  sensitive = true,
  keywords={algorithm, input, output, return, if, else, for, while},
  numberstyle=\footnotesize,
  numbersep=9pt,
  showstringspaces=false,
  breaklines=true,
  comment=[l]{//},
  morecomment=[s]{/*}{*/},
}

\lstdefinestyle{grammar}{
  belowcaptionskip=1\baselineskip,
  basicstyle=\rmfamily\mdseries\footnotesize,
  breaklines=true,
  language=alg,
  frame=lines,
  xleftmargin=\parindent,
  showstringspaces=false,
  mathescape=true,
  numberstyle=\tiny,
  keywordstyle=\bfseries
}

\lstdefinestyle{pseudocode}{
  belowcaptionskip=1\baselineskip,
  breaklines=true,
  frame=leftline,
  language=alg,
  xleftmargin=\parindent,
  showstringspaces=false,
  mathescape=true,
  numberstyle=\tiny,
  keywordstyle=\bfseries
}

\lstdefinestyle{examples}{
  belowcaptionskip=1\baselineskip,
  breaklines=true,
  frame=shadowbox,
  language=C,
  xleftmargin=\parindent,
  showstringspaces=false,
  mathescape=true,
  numberstyle=\tiny,
}

\begin{document}

\title{\toolname: Syntax-Guided Synthesis 
of \\Expressions with Arrays and Quantifiers}

\author{Elizabeth Polgreen\inst{1,2} \and
Sanjit A. Seshia\inst{1}}
\authorrunning{E. Polgreen et al.}
\institute{UC Berkeley \and University of Edinburgh}

\maketitle              % typeset the header of the contribution

\begin{abstract}%4 sentences
% State the problem
%
Program synthesis is the task of automatically generating expressions that
satisfy a given specification. Program synthesis techniques have
been used to automate the generation of loop invariants in code, synthesize
function summaries, and to assist programmers via program sketching. 
Syntax-guided synthesis has been a successful paradigm in this area, however,
one area where the state-of-the-art solvers fall-down is reasoning about
potentially unbounded data structures such as arrays where both specifications
and solutions may require quantifiers and quantifier alternations.
% Proving properties of systems frequently requires
% the user to provide hand-written invariants and pre- and post-conditions. 
% A~significant body of work exists attempting to automate the generation
% of loop invariants in code, but the state of the art cannot yet tackle
% the combination of quantifiers and potentially unbounded data structures.
%
We present \toolname, a synthesis algorithm based on restricting the 
synthesis problem to generate
candidate solutions with quantification over a finite domain, and then
generalizing these candidate solutions to the unrestricted domain of the
original specification.
We report experiments on invariant synthesis benchmarks and on program sketching 
benchmarks taken from the Java StringUtils class and show that our technique
can synthesize expressions out of reach of all existing solvers.

% Our
% algorithm is, in principle, able to synthesize predicates with arbitrarily
% many levels of alternating quantification.  We report experiments that
% require invariants with one alternation that are already out of reach of all
% existing solvers.
%
\end{abstract}

\section{Introduction}

%program synthesis int3o
A program synthesis solver aims to find a program or expression $P$ 
that reasons about a set of input arguments $x$ and satisfies some logical
specification $\sigma$ for all possible values that those input arguments may take.
That is, synthesis solvers find a solution to formulas of the 
following form:
$$
\exists P \forall \vec{x}\,\, \sigma(P,\vec{x}).
$$
However, the application of program synthesis to many real world-domains can be limited
by the abilities of state-of-the-art solvers to reason about input arguments
of types that are
typically found in these domains. One instance of this is data-structures
like arrays, which, in many software applications, 
may be so large as to be practically 
considered infinite.
The size of these data-structures introduces the need
for additional quantification, beyond that required for 
standard program synthesis, in both specifications and solution expressions.

%now invariants
\paragraph{Invariant synthesis: }Consider the example of invariant synthesis. 
Symbolic methods for proving properties of programs use inductive arguments
to reason about execution traces of potentially unbounded length. 
Many programs or software systems have a potentially unbounded
state space, which might arise from unbounded data structures in memory or
an unbounded number of threads or an unbounded numbers of machines on a
network.  Inductive arguments for non-trivial properties of such systems
require quantification over the state-holding elements, e.g., the elements
of a potentially unbounded array, i.e., we require invariants of the form
$$\forall i \in \{0, \ldots, n\}. I(i)$$
where $n$ denotes the number of components and where $I(i)$ is a property of
the state of the component with index $i$. 
% The community has invested a large amount of effort into this problem,
% and identified a broad variety of special cases in which reachability
% properties of parametric systems are decidable. For the unrestricted
% case, the community has devised numerous heuristic methods for guessing
% and possibly refining the 
% predicate~$I$~\cite{DBLP:conf/aplas/KongJDWY10, DBLP:conf/kbse/NguyenDV17,
% DBLP:conf/nips/SiDRNS18}. 
%
%
This paper addresses the general case of program synthesis in which we wish to 
synthesize an expression that may reason about unbounded data-structures using
quantifiers, and which satisfies some logical specification $\sigma$ which also
may contain additional quantifiers. 
Both the expression and the specification may contain
quantifier alternations. 
Consider the task of synthesizing
a safety invariant for the loop shown in Figure~\ref{fig:alternating_q}.
The two arrays $A$ and $B$
are initially equal.  The elements of $A$ are swapped nondeterministically.
We use~$\ast$~to indicate a non-deterministic choice. 
The assertion in this snippet checks that for every element of $A$
there exists an equal element in~$B$.  The most natural way to formalize
this property is to use a formula with one quantifier alternation.  
% The
% given property is inductive as-is, and it is easy for state-of-the-art
% SMT solvers to show validity of the corresponding verification conditions.

\begin{wrapfigure}{l}{0.5\linewidth}
% \begin{figure}
% \center
\begin{lstlisting}[style=examples]
int $A[\,]$; int $B[\,]$; int c;
assume: $\forall i \,\,. A[i]==B[i]$
assume: $c>0$

while($\ast$)
  x'=$\ast$; y'=$\ast$;
  $A'[x]=A[y]$
  $A'[y]=A[x]$
  swap($A[x]$, $A[y]$)
  $\forall i\, A'[i]=A[i]+c \wedge B'[i]=B[i]+c$

assert: $\forall x\,\exists y \,\,.A[x]=B[y]$
\end{lstlisting}
\caption{A safety
property with a quantifier alternation. \label{fig:alternating_q}}
% that is naturally specified using one quantifier alternation}

% \end{figure}
\end{wrapfigure}

There are many verification problems like the one above, in many forms
and shapes.  In this particular instance, while it is easy to verify that
the assertion above holds, it is difficult for humans to write such an invariant, 
and thus,
there is a need for algorithmic methods that identify these invariants
automatically.  Nevertheless, to the best of our knowledge, no existing
%
%syntax-guided 
%
synthesis engine is able to generate the inductive invariant for the example
above. When formulated as a synthesis problem, 
CVC4~\cite{DBLP:journals/corr/abs-1806-08775} version 1.8 returns ``unknown'',
and when formulated as a Constrained Horn Clause problem
Z3~\cite{DBLP:conf/tacas/MouraB08,DBLP:journals/fmsd/KomuravelliGC16} (version 4.8.7)
also returns
``unknown''.

\paragraph{Sketching: }Another example application of program synthesis is 
program sketching~\cite{DBLP:conf/asplos/Solar-LezamaTBSS06}, 
in which a programmer provides
a sketch of a program with some specification and asks a synthesizer to fill the 
the holes. The idea behind sketching is that it is often easier to write a specification
that describes what some block of code should do than it is to write the code itself. 
However, writing specifications for programs that reason about arrays often 
requires quantification that is not currently supported by state-of-the-art solvers.
This motivates the need for additional quantification support in specifications.
To illustrate this point, we compile a set of program sketching benchmarks that require a program synthesizer
to synthesize fragments of the Java StringUtils class, one of the most
used Java Utility Classes. A string in Java, or C, is
represented by an array of chars, and we model this as an array of integers. In Figure~\ref{fig:sketch_ex}
we show an example program sketch asking the synthesizer to synthesize part of 
a method, denoted by $??$, which returns true if a string contains a given character. 

\begin{wrapfigure}{r}{0.6\linewidth}
\begin{lstlisting}[style=examples]
int str[];
int strLen;
int ch;

bool contains(str, strLen, ch)
{
  for(int i=0; i<strLen, i++)
    if(??)
      return true;

  return false;    
}

assert($\exists$i. str[i]=ch
  $\iff$contains(str, strLen, ch))
\end{lstlisting}
\caption{Sketch example\label{fig:sketch_ex}}
\end{wrapfigure}

The natural way to write a specification for this code fragment, which should
return $true$ if an array of chars contains a target char, and $false$ otherwise, is to
use an existential quantifier:
$(\exists i \, str[i] = ch)\iff contains(i, str[i])$.
Such quantification is not permitted within the syntax-guided synthesis
format~\cite{sygus}, nor can such a problem be formulated 
within the constrained-horn-clause format~\cite{chc}. The existential
quantifier puts this problem out of reach of state-of-the-art 
Syntax-Guided Synthesis solver
CVC4~\cite{DBLP:journals/corr/abs-1806-08775}.

\paragraph{\toolname:} 
In this paper we present \toolname: Synthesis via Restriction and
Generalization.  
The algorithm is based around a CounterExample Guided Inductive
Synthesis~(CEGIS)~\cite{DBLP:conf/asplos/Solar-LezamaTBSS06} solver, a well-known
paradigm for solving program synthesis problems of the form $\exists P
\,\forall \vec{x}\, \sigma(P,\vec{x})$.
%
% where $\sigma$ and $P$ are the specification and
%the program to be synthesized.  
Current state-of-the-art techniques can only
handle quantifier-free specifications, $\sigma$, and programs,$P$.  
The original CEGIS algorithm
performs synthesis in two stages: first it synthesizes a candidate solution
that works for a \emph{simpler} version of the specification (i.e., it works for a
subset of the possible inputs) and then verifies whether that candidate
solution works for the full specification (i.e., for all the possible
inputs).  Using this principle, we perform synthesis by restricting a
specification that contains quantifiers over infinite domains into a \emph{simpler}
quantifier-free specification on a restricted domain.  We synthesize candidate
solutions for this simpler specification, and then attempt to generalize
them to satisfy the full specification.  \toolname integrates well with
existing syntax-guided synthesis solvers: by reducing complex synthesis
specifications with nested quantification over infinite domains to
quantifier-free specifications over restricted domains, we take advantage of
state-of-the-art synthesis solvers for quantifier-free theories.

We present an instance
of the algorithm that solves formulas with quantifiers over the indices of
arrays. In order to evaluate this algorithm,since there are currently
no benchmarks for Syntax-Guided Synthesis that use arrays and quantifiers, we compile a set 
of benchmarks in the standard Syntax-Guided Synthesis Interchange Format~\cite{sygus}.
This set of benchmarks contains
basic invariant synthesis tasks, benchmarks adapted from the software
verification competition~\cite{10.1007/978-3-030-45237-7_21}, and 
program sketching examples which require the synthesis solver to synthesize
parts of the Java StringUtils class.
We observe that the algorithm outperforms the state-of-the-art solvers; we
hypothesize that this is owed to the fact the array theory employs the
\emph{small model property}~\cite{DBLP:conf/vmcai/BradleyMS06}, i.e., the
validity of quantifier-free array formulas can be determined in a finite
domain by computing a set of indices that is often surprisingly small. 

We conjecture that some variant of this algorithmic framework 
would be extensible to any
theory over infinite domains if they have the small model property.  That
is, for any formula $f$ over an infinite domain, where a formula $f_b$ can
be constructed over some finite domain and $f_b$ is satisfiable iff $f$ is
satisfiable.

\paragraph{Contributions:  }The contributions of this paper are:
\begin{itemize}
\item \toolname:
a general program synthesis algorithm which can synthesize expressions
containing quantifiers, which satisfy specifications containing quantifiers. 
As far as we know, this is the first solver general enough to automatically
synthesize expressions containing alternating quantifiers. 
\item a new set of program synthesis benchmarks based on the Java StringUtils class 
expressed in
 Syntax-Guided Synthesis Interchange-Format (SyGuS-IF)~\cite{sygus} with additional quantification,
 basic invariant synthesis tasks and SV-comp~\cite{10.1007/978-3-030-45237-7_21} benchmarks.
\end{itemize}
\begin{figure*}
    \centering
\input{new_alg.tikz}
    \caption{SynRG: algorithm for synthesis of programs with quantifiers and arrays}
    \label{fig:basic_alg}
  \end{figure*}
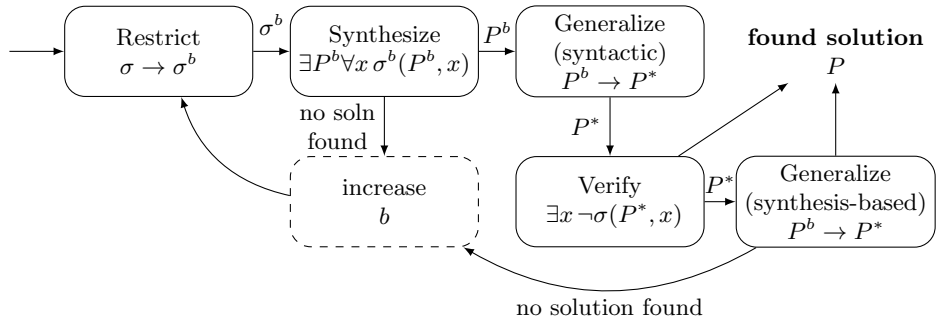
\section{Background}
\label{sec:prelim}
\subsection{Program synthesis}
Program synthesis can be framed as the following existential second-order logic formula:
$$
\exists P .\, \forall \vec{x}.\, \sigma(\vec{x}, P)$$
where $P$ ranges over functions
(where a function is represented by the program computing it)
and $\vec{x}$ ranges over ground terms. 
We interpret the ground terms over some domain, and we use ${\cal I}$ to denote
the set of all possible inputs to the function, and so in the formula above $\vec{x} \in {\cal I}$. Program sketching is a specific instance of this problem, where the specification constitutes a sketch of some program that contains holes and some assertions about the behavior of that code, and the holes are expressions to be synthesized. 
We allow $P$ and $\sigma$ to contain linear integer arithmetic, arrays and quantification over array indices.  We restrict the array indices to terms in Presburger arithmetic. 
% In order to explain our exploitation of the small-model property of arrays, we restrict the formula allowed in $\sigma$ to fall in a subset of $\exists^*\forall^*_{\mathbb{Z}}$ of arrays, where the subscript on $\forall$ indicates that the quantifier is only over array index variables. 

CEGIS~\cite{DBLP:conf/asplos/Solar-LezamaTBSS06} is an algorithmic framework often used
 to tackle the program synthesis problem described above.
 It consists of two phases;
the synthesis phase and the verification phase: Given the specification of the
desired 
program, $\sigma$, the inductive synthesis procedure produces a candidate program~$P$ 
that satisfies the specification $\sigma(\vec{x},P)$ for all $\vec{x}$ in a set of inputs ${\cal I_G}$ 
(which is a subset of ${\cal I}$). The candidate program $P$ is passed to the verification phase,
which checks whether $P$ 
is a full solution, i.e., it satisfies 
the specification $\sigma(\vec{x}, P)$ for all $\vec{x}$. If so, we have successfully 
synthesized a full solution and the algorithm terminates. Otherwise, the verifier passes a counterexample, i.e., the satisfying assignment to $\vec{x}$, which is added to the set of inputs ${\cal I_G}$ passed to the synthesizer, and the 
loop repeats.

\subsubsection{Safety invariants}
The synthesis of safety invariants can be formulated as a general program synthesis problem, and the Syntax Guided Synthesis Competition~\cite{DBLP:journals/corr/abs-1711-11438} has a track dedicated to this. We use safety invariants as an exemplar of program synthesis in this paper although our algorithm is able to tackle more general synthesis problems. Given a loop with a loop variable $x$, some initial conditions $init(x)$, a transition relation $trans(x,x')$, and a post condition $post(x)$, we synthesize an invariant $inv$ such that 

\begin{align*}
\exists inv(x) \forall x\, \\
init(x) &\implies inv(x) \wedge \\
inv(x) \wedge trans(x,x') &\implies inv(x') \wedge \\
inv(x) &\implies post(x). \\
\end{align*}

Finding the simplest loop invariants involves solving this formula with a single quantifier alternation. 
This paper extends this to permit quantifiers in specifications as well as the synthesized invariants. Use cases that benefit from
specifications containing quantifiers 
quantification are nested loops, the initialization of arrays, or
asserting properties about arrays. 
 Consider the example given in
Figure~\ref{fig:running_ex}.
A reachability problem such as this one requires reasoning with alternating
quantification: a candidate inductive invariant for this loop must satisfy
the base case of the induction proof, i.e., that
$\forall \vec{x}\,\, (init(\vec{x}) \implies inv(\vec{x})),$
where $\vec{x}$ denotes the set of all possible inputs to the program, 
$init(\vec{x})$ asserts that the initial conditions of the program hold, and
$inv(\vec{x})$ is the inductive invariant we wish to synthesize.  This is
equivalent to $$\forall \vec{x}\,\, (\neg init(\vec{x}) \vee inv(\vec{x})).$$  
For Figure~\ref{fig:running_ex}, 
the initial conditions are 
$(c<0) \, \wedge \forall i \,A[i]\geq 0$, and thus we obtain the following base case
of the induction proof:
$$\forall A,c \,\, (c < 0 \, \vee \,(\exists i \,\,A[i])<0) \, \vee \,inv(A,c).$$
Note the quantifier alternation in $\forall A,c \exists i$.  Nested quantifiers
are not supported by recent works in synthesis of quantified
invariants~\cite{DBLP:conf/atva/GurfinkelSV18,
DBLP:conf/cav/FedyukovichPMG19}, and are explicitly prohibited in the
Constrained-Horn Clause~\cite{chc} and in the SyGuS
competition~\cite{sygus} formats.  

% Consequently, the simple inductive
% invariant for this example cannot be synthesized by the leading solvers in these competition:
% CVC4~\cite{DBLP:conf/cav/ReynoldsBNBT19} version 1.9\,[pre-release]; the
% Z3~\cite{DBLP:journals/fmsd/KomuravelliGC16} Horn solver version
% 4.8.7; FreqHorn~\cite{DBLP:conf/cav/FedyukovichPMG19} or
% QUIC3~\cite{DBLP:conf/atva/GurfinkelSV18}.
\begin{figure}
\begin{center}
\begin{lstlisting}[style=examples]
int $A[\,]$;
int $c=\ast$;
assume: $c>0$
assume: $\forall i \,\,. A[i]\geq 0$
  
while($\ast$)
    $\forall i\,\, A'[i]=A[i]+c$

assert: $\neg \exists i\, A[i]<0$
\end{lstlisting}
\caption{Simple running example\label{fig:running_ex}}
\end{center}
\end{figure}

\section{Algorithm overview}

The classic CEGIS algorithm breaks down a hard problem into an easier
problem for synthesis, and then attempts to generalize a solution to the
easier problem to the full problem.  That is, the synthesis phase
attempts to synthesize a solution that works only for a subset of inputs,
i.e., it solves $\exists P \forall \vec{x} \in {\cal I_G} \sigma(P, \vec{x})$.  The
verification phase then checks whether a candidate solution 
satisfies the specification for all possible inputs.

We apply a similar approach to CEGIS:
We take a specification
$\sigma$, which contains quantification over the infinite domain of arrays.  We restrict
the domain of arrays in $\sigma$, generating a restricted-domain specification $\sigma^b$, 
which considers only $b$ elements of each array.  
This
specification then only contains quantification over finite domains, which can be 
removed with exhaustive quantifier instantiation. 
We synthesize a solution to $\sigma^b$ with an existing state-of-the-art
Syntax-Guided Synthesis solver.  We use $P^b$ to denote a solution to
$\exists P \forall \vec{x} \sigma^b(P, \vec{x})$.  We then attempt to generalize $P^b$,
to give a candidate solution to the full specification, which we denote
$P^*$, and verify whether $P^*$ is a solution to $\sigma$, i.e.,
$\exists x \,\, \neg\sigma(P^*,\vec{x})$. The generalization has two phases: a syntactic
generalization phase and a synthesis-based generalization phase. If we fail to find a solution, we 
increase the
bound $b$ used for generating $\sigma^b$.
An overview of this general algorithmic framework is illustrated in Figure~\ref{fig:basic_alg}.

\paragraph{\textbf{Verification phase: }}
The verification phase, using a standard SMT-solver, guarantees the soundness of our approach.
There are two verification phases; the first verifies the candidate produced by
the first (syntactic) generalization phase. If this verification fails, then 
the incorrect solution is used by the synthesis-based generalization phase.
The second verification phase is embedded inside the synthesis-based generalization,
and if the synthesis-based generalization phase fails to produce a candidate that passes
verification, we relax the restriction on
% We verify that the candidate program $P$ satisfies the specification
% $\sigma$ using an SMT-solver. This guarantees soundness of our approach. 
the array size, i.e., the arrays in $\sigma^b$
increase in size, but are still smaller than the arrays in $\sigma$ 
% \footnote{However, as illustrated by the dashed line in Figure~\ref{fig:basic_alg}, it may be interesting to use the counterexamples to block specific restricted-domain
% candidate solutions, and repeat the synthesis step with the same restricted-domain specification.}.

\paragraph{\textbf{Synthesis phase: }}
The synthesis phase solves the formula $\exists P^b\, \forall \vec{x}\,
\sigma^b(P^b, \vec{x})$.  This formula contains only theories permitted in
SyGuS-IF~\cite{sygus}, the universal input format for SyGuS solvers, and 
we are able to apply standard synthesis-guided synthesis algorithms.
The synthesis problem is not, in general, decidable, and the syntax-guided synthesis
algorithms that complete this task are not complete~\cite{DBLP:journals/corr/CaulfieldRST15}, and consequently
 \toolname is also not complete.
The synthesis phase initially constructs the bounded synthesis query without
using a syntactic template, and calls a solver with a time-out of $2$s. If this query returns a
solution then we proceed to the generalization phase. If it does not, we add a syntactic template
to the synthesis query and call a solver with a time-out of $60$s. These time-outs are heuristics, 
and increasing the time-out may allow \toolname to solve more benchmarks, but generally slower since each iteration takes longer. 

The syntactic template comprises a non-terminal for each non-array parameter or return type,
 all non-array parameters, all operators in linear integer arithmetic and array select operators for indices up to the bound of $\sigma_b$.
 For example, this syntactic template would be used for a function with two array parameters, two integer parameters and a boolean return type, and a bound of $2$:
\begin{lstlisting}[style=grammar]
Program::= B
B::= B $\wedge$ B|B $\vee$ B| not B|I $\geq$ I|I $\leq$ I|I = I
I::= 0|1|y|z|I-I|I+I|arr1[0]|arr1[1]|arr2[0]|arr2[1]
\end{lstlisting}
Although these synthesis queries are executed sequentially, they are not dependent one each other so they could practically be executed in parallel.

% \emph{Running example: }the
% restricted-domain specification for the problem shown in Figure~\ref{fig:running_ex} is Equation~\ref{eq:bounded_spec}. 
% A~possible solution to this restricted-domain specification is:
% %
% $
% A[0]\geq 0 \wedge A[1]\geq 0 \wedge c \geq 0\,.
% $

%We use the notation $\sigma^b$ to denote the simplified specification and $P^b$ to denote a program $P$ which satisfies $\sigma^b(P, \vec{x})$ for all $x$. 
%
% Given an input specification $\sigma$, our algorithm iterates through the following phases:
% \begin{itemize}[align=left]
%   \item Restriction:  Derive a restricted-domain specification $\sigma^b$ such that $\sigma^b(P, \vec{x})$ and $\sigma(P, \vec{x})$ are equisatisfiable. 
%   \item Synthesis:  Synthesize a program $P^b$ which satisfies $\sigma^b$, i.e., find a satisfying assignment to the formula $\exists P^b \forall \vec{x}\,\, \sigma^b(P^b, \vec{x})$.
%   \item Generalization: Attempt to generalize $P^b$, generating a candidate $P^*$ which may be a solution to $\sigma$. This generalization approximately reverses the transformation performed in step 1.
%   \item Verification: Check whether $P^*$ is a solution to $\sigma$. If it is not, we increase the size of the small model used in step 1. 
% \end{itemize}
\paragraph{\textbf{Restriction and Generalization: }}In the following sections, we discuss the fragments of array logic for which such a restricted-domain specification 
is guaranteed to exist. 
We then give details of a specific instance of this algorithmic framework, where the restricted-domain specification
considers explicitly $b$ elements of every array, and the generalization approach is a two staged approach based on applying a series of syntactic rules and then, if that fails, an additional synthesis-based generalization procedure.

\section{The small model property}
\label{sec:small-model}
\label{sec:decidability}

% Many techniques for verification of parametric systems depend on the small model 
% theorems~\cite{DBLP:conf/vmcai/BradleyMS06,DBLP:conf/sp/FranklinCDS10}, which allow
% the procedure to infer satisfiability of an infinite model from a finite model. 

The algorithm we have presented benefits from the existence of a restricted-domain program $P^b$ that can be generalized to the unrestricted domain. In this section we prove that such a restricted-domain program is guaranteed to exist for a restricted fragment of array theory. 

\subsection{Fragment of array theory}
Consider the case where the specification $\sigma$ and the solution to be synthesized $P$ are restricted in such a way that the verification query can be written as a boolean combination of array properties~\cite{DBLP:conf/vmcai/BradleyMS06}. 
\begin{definition}{Array property: }
\label{def:frag}
an array theory formula is called an array property~\cite{DBLP:conf/vmcai/BradleyMS06,DBLP:series/txtcs/KroeningS16} \emph{iff} it is of the form
$$ \forall i_1\ldots \forall i_k \in T_I, \phi_I(i_1, ... i_k) \implies \phi_V(i_1, ..., i_k),$$
where $T_I$ is the index theory, $i_1, ..., i_k$ are array index variables, $\phi_I$ is the index guard and $\phi_V$ is the value constraint, and the formula also satisfies the following conditions:
\end{definition}
\begin{enumerate}
  \item the index guard $\phi_I$ must follow the grammar
\begin{lstlisting}[style=grammar,language=alg] 
iguard::= iguard$\wedge$iguard|iguard$\vee$iguard|iterm$\le$iterm|iterm$=$iterm 
 iterm::= $i_1$|...|$i_k$|terms
  term::= integer_constant|integer_constant$*$index_identifier|term$+$term
\end{lstlisting}
The {\rmfamily\mdseries\footnotesize index\_identifier} used in {\rmfamily\mdseries\footnotesize term} must not be one of the universally quantified variables.
\item The index variables $i_1, ..., i_k$ can only be used in array read expressions.
\end{enumerate}

If this is the case, then we know that the verification query is solvable~\cite{DBLP:conf/vmcai/BradleyMS06} and, crucially, that there is a finite set of index terms such that instantiating universally quantified index variables from only this set is sufficient for completeness. This set of index terms, $\mathcal{R}$ is made up of all expressions used as an array index term, or inside index guards, in $\phi$ that are not quantified variables.

The example shown in Figure~\ref{fig:decidable}, shows an invariant synthesis problem with one possible target solution. The verification query for checking the inductive step of this target solution is:
\begin{align*}
\forall x (x < i \implies a[x]=0 ) \\ 
\wedge\, a'[i]=0 \wedge \forall j\neq i. a'[j]=a[j] \\
\wedge\, \exists x .(x \leq i \wedge a'[x]\neq 0).
\end{align*}
We instantiate the existential quantifier with a fresh variable $z$:
\begin{align*}
\forall x (x < i \implies a[x]=0 ) \\ 
\wedge\, a'[i]=0 \wedge \forall j\neq i. a'[j]=a[j] \\
\wedge\, (z \leq i \wedge a'[z]\neq 0).
\end{align*}
The set of index terms $\mathcal{R}$ is $\{i, z\}$. If we replace the universal quantifier $\forall i P(i)$ with a conjunction $\bigwedge_{i \in \mathcal{R}} P(i)$, we get the following quantifier-free formula:
\begin{align*}
(z < i \implies a[z]=0 ) \\ 
\wedge\, (a'[i]=0) \wedge  (z\neq i \implies a'[z]=a[z]) \\
\wedge\, (z \leq i \wedge a'[z]\neq 0).
\end{align*}

Thus, it is sufficient to verify the candidate $P$ by only considering two elements of the array, $a[z]$ and $a[i]$, provided we consider the cases where $z<i$, $z=i$, and $z>i$. There is a restricted-domain candidate program $P^b$ and a restricted-domain specification $\sigma^b$ such that $\exists x \neg \sigma^b(P^b,x)$ is equisatisfiable with the original verification formula. In this case the restricted-domain specification for the inductive step would be $(P^b(a) \wedge a'[i]=0 \wedge (z\neq i \implies a'[z]=a[z])) \implies P^b(a')$ and the restricted-domain program $P^b(a)$ is $z<i \implies a[z]=0)$. Given that we only need to reason about array indices $a[z]$ and $a[i]$, if we find a solution that works for an arrays of size $2$, we will have a solution that we can generalize to the infinite case by a procedure detailed in Section~\ref{sec:generalize} that is based on reversing the steps we used to obtain the restricted-domain specification.

However, without knowing the solution $P$ in advance, we cannot determine the size of the set $I$ needed for the verification query. Consequently, we use a heuristic approach in this work where we begin with two array elements and increase the number of elements if we are unable to find a solution. The exact process we use is detailed in Section~\ref{sec:restriction}.
We note that the restricted-domain synthesis query itself does not fall within a decidable fragment~\cite{DBLP:journals/corr/CaulfieldRST15}, even for this array fragment, and so, perhaps unsurprisingly, the unrestricted-domain synthesis problem is in general undecidable.  

%TODO replace with example 1? that only has a model of size 1 though

\begin{figure}
\begin{lstlisting}[style=examples,linewidth=5cm]
int $A[\,]$;
int $i=0$;
  
while(i<50)
  A'[i]=0; i'=i+1;
  invariant: $\forall x, (x<i) \implies A[x]=0$

assert: $x<50 \implies A[x]=0$
\end{lstlisting}
\caption{Safety invariant expressed in array property fragment\label{fig:decidable}}
\end{figure}

\subsection{Beyond the array property fragment}
The array property fragment in Definition~\ref{def:frag} is restrictive, but expressive enough for many 
benchmarks we adapted from the SV-Comp~\cite{10.1007/978-3-030-45237-7_21}. 
% \begin{itemize}
% \item 
%
As an exemplar of a synthesis problem that falls outside this fragment, consider that we have synthesized an invariant that states that
array $A$ contains at least one element that is not in array $B$. That is 
$\exists i \forall $j$ A[i] \neq B[j]$. The verification of this invariant will include the subformula $\exists $A,B$\, \forall i \exists j. A[i]=B[j]$. This formula falls outside the decidable fragment described above, and formulas of the form $\exists x \forall i \exists j$, where $x$ is some array and $i$ and $j$ are indices, are in general undecidable. This can be shown by reduction to the termination of integer loops~\cite{DBLP:conf/vmcai/BradleyMS06}. Consequently we are unable to show that a finite model exists. However, our experimental evaluation shows that the approach we present in this paper is a heuristic that can be used to some problems that fall outside of the decidable fragment.

\section{Restriction of $\sigma$ to $\sigma^b$}
\label{sec:restriction}
We aim to generate a 
modified specification $\sigma^b$ that considers a 
finite set of $b$ index terms for each array. 
We do this by bounding the 
length of the arrays in the specification to length $b$,
 by replacing any predicate $e$ in $\sigma$ that 
reasons about an array index $i$, with an implication $(0\leq i < b) \implies e$. 
% The restricted-domain specification is guaranteed to always be weaker than the original specification, i.e., 
% it permits more solutions than the original specification, and the sequence of restrictions we build
% as we increase $b$ is monotonic. 

\begin{algorithm}
  \label{alg:bound-arrays}
\KwData{$\sigma$, bound b}
\KwResult{$\sigma^b$} 
idx: list of array indices \;
Qidx: list of quantified array indices\;
$\sigma^b\leftarrow \emptyset$\;
idx$\leftarrow \emptyset$\;
Qidx$\leftarrow\emptyset$\;
\For{constraint  $c \in\sigma$}{
    $c^b \leftarrow$ boundQuantification($c$, b, idx, Qidx)\;
    $c^b \leftarrow$\{(idx $< b$ )$\implies c^b$\}\;
    idx$\leftarrow \emptyset$\;
    $\sigma^b \leftarrow c^b$\;
}
  \Return $\sigma^b$
  \caption{Pseudocode for generating $\sigma^b$}
\end{algorithm}

 \begin{algorithm}
 \KwData{expression $e$, bound $b$, idx, Qidx}
 \KwResult{finite-domain expression $e^b$, updated idx and Qidx}
 \If{$e$ is $\forall$ or $e$ is $\exists$}
 {
  idx $\leftarrow$ Qidx\;
  Qidx $\leftarrow \emptyset$\;
 }
 \For{operand o $\in$ e.operands} {
  $o \leftarrow$boundQuantification(o, b, idx, Qidx)\;
 }
 \uIf{($e $ is $ \forall$  or $e$ is $\exists$) $\wedge$ Qidx $\neq \emptyset$ }
 {
   $e.Predicate\leftarrow \{($Qidx$<B)\implies e.Predicate$\}\;
   Qidx$\leftarrow \emptyset$ \;
 }
 \uElseIf{$e$ is an array element}
 {
  Qidx $\leftarrow$getIndex($e$)\;
 }
 \Return $e$\;

 \caption{boundQuantification: Algorithm for bounding an expression}
 \end{algorithm}

The algorithm for bounding arrays, shown in Algorithm~\ref{alg:bound-arrays}, 
applies these rules recursively on the syntax tree of each constraint. 
This method of considering only the first $b$ elements of the arrays
may require us to consider a larger specification than strictly necessary. For instance, 
suppose we have a specification that reasons only about array element $x[99]$. 
Our algorithm will not work until we have increased $b$ to $100$, and we then have to solve a synthesis query that
considers all elements $x[0]..x[99]$. 
Future work will explore more sophisticated heuristics for generating this specification.

\paragraph{Remove quantification: } 
Once we have obtained this restricted-domain specification for $\sigma$, all 
quantification is now over finite domains. We can hence
 use exhaustive quantifier instantiation to remove all quantifiers over array indices 
 and replace universal quantifiers with conjunctions and existential quantifiers with disjunctions. 
This exhaustive quantifier instantiation is possible only because we have bound the size of the arrays to make the data types finite. 

% We apply the following rules exhaustively:

% $$\frac{\psi[\forall \vec{i} (\varphi_I(\vec{i}) \implies \varphi_V(\vec{i}))]}
% {\psi[\bigwedge_{0 \leq i < b} (\varphi_I(i) \implies \varphi_V(i))]}$$

% $$\frac{\psi[\exists \vec{i} (\varphi_I(\vec{i}) \implies \varphi_V(\vec{i}))]}
% {\psi[\bigvee_{0 \leq i < b} (\varphi_I(i) \implies \varphi_V(i))]}$$
\paragraph{Running example:}
% \paragraph{Running example: } 
Consider the example presented in Figure~\ref{fig:running_ex}. 
The constraint asserting that the invariant holds at the initial conditions would be:
\begin{align*}
(c>0) \wedge ( \forall i \,\,. A[i]\geq 0) \implies inv(c,A) \,\,\bigwedge\\
 inv(c,A) \wedge (\forall i\,\, A'[i]=A[i]+c) \implies inv(c,A') \,\,\bigwedge\\
 inv(c,A) \implies \neg \exists i\, A[i]<0.
\end{align*}
If we apply these steps to our running example, bounding the arrays to size two, we get the following constraints:
\begin{align}
\begin{split}
(c>0) \wedge ( \bigwedge_{0 \leq i<2} A[i]\geq 0) \implies inv(c,A) \,\, \bigwedge\\
inv(c,A) \wedge (\bigwedge_{0 \leq i < 2} A[i]=A[i]+c)\implies inv(c,A')\,\,\bigwedge\\
inv(c,A) \implies \neg \bigvee_{0 \leq i < 2}A[i]<0 \label{eq:bounded_spec}
\end{split}
\end{align}

\begin{algorithm}
\KwData{expression $e$, bound $b$}
\KwResult{quantifier-free expression}
 \For{operand $o \in e.operands$}
{
  removeQuantifers($o$, $b$)\;
}

\If{$e$ is $\forall$ or $e$ is $\exists$}
{
  $v \leftarrow e.Binding$ \;
  $P \leftarrow e.Predicate$ \;

  \uIf{$e$ is $\forall$}
  {
  $e_{qf} \leftarrow$ conjunction\;
  }
  \Else
  {
  $e_{qf} \leftarrow$ disjunction\;
  }

  \For{$0 \leq i < b$}
  {
     $P_i \leftarrow $  replaceVarWithConstant($P$, $v$, $i$)\;
\tcc{add $P_i$ to the operands of $e_{qf}$, which is a conjunction or disjunction.}
     $e_{qf}.operands \leftarrow P_i$ \;
  }
  \Return $e_{qf}$
}
\Return $e$
\caption{removeQuantifiers: This psuedocode is simplified to 
handle only quantifiers that bind only to a single variable}
\end{algorithm}

% \paragraph{Cast to integers}
% In some (limited) cases, it might be easier to find a solution for a problem using integers and LIA instead of bitvectors,
% due to the non-linear nature of bitvectors, and so we also provide this transformation. 
% \begin{itemize}
%   \item find all bitvector types (variables and constants) and replace with integers
%   \item find all operations specific to bitvectors that can be easily converted to integers, and convert to integers
%   \item if we find bitvector operations that cannot be represented in LIA (e.g., shift), abort. 
% \end{itemize}

\section{Generalization}
\label{sec:generalize}
Assuming the synthesis block has found a solution $P^b$, we now attempt to
generalize this solution to obtain a candidate $P^*$ that may satisfy
the full specification, by introducing quantifiers in the place of conjunctions or disjunctions.
The generalization has two phases: a syntax-based quantifier introduction procedure, which, if it
fails to produce a solution that works for the full array, is then used to provide syntactic-guidance
to a synthesis-based generalization phase.
\begin{algorithm}
\KwData{finite-domain expression $e$}
\KwResult{An unrestricted-domain expression}
\For{operand $o \in e.operands$}
{
  generalize($o$)\;
}
\tcc{set of matching operands}
$Ops \leftarrow \emptyset $\; 
\tcc{set of sets of matching operands}
$Sets \leftarrow \emptyset $\; 

\If{$e$ is $\wedge$ or $\vee$}
{
    Ops$\leftarrow${e.operand[0]}\;
    Sets$\leftarrow$Ops\;
    N$\leftarrow$e.operands.size()\;

    \For {$1\leq i < N$}
    { 
      placed$\leftarrow$false
      \For {set$\in$Sets}
      {
        \If{compareExpr(set, e.operand[i])}
        {
          set$\leftarrow$E.operand[i]\;
          placed$\leftarrow$true\;
        }
      }

      \If{!placed}
      {
       newSet$\leftarrow$e.operand[i]
       Sets$\leftarrow$newSet 
      }
    }     

    result$\leftarrow$true;   
    \For{set$\in$Sets}
    {
    \tcc{Replace array indices with local variables}
      P$\leftarrow$replaceIndicesWithVars(set, $v_{loc}$)\;
      \uIf{$e$ is $\wedge$}
      {
        quantifiedExpr$\leftarrow \forall v_{loc} \,\,P$ 
      }
      \uElseIf{$e$ is $\vee$}
      {
        quantifiedExpr$\leftarrow \exists v_{loc} \,\,P$
      }
    result$\leftarrow$result$\wedge$quantifiedExpr
    }

    \Return result;  
}
\caption{syntactic generalize: algorithm for reintroducing quantifiers}
\label{alg:reintroduce_quantifiers}
\end{algorithm}
\subsection{Syntactic-generalization}
We implement a syntax-based quantifier introduction procedure based on
identifying conjunctions or disjunctions of predicates that use array
indices. We describe the steps for universal quantifiers, and note that existential
quantifiers can be introduced by treating disjunctions in the same way. In order to introduce a universal quantifier, in place of an expression:
\begin{itemize}
  \item the expression must be a conjunction of predicates that reason about array elements;
  \item replacing an array index in the same location in each predicate with a new variable must result in equisatisfiable predicates; 
  \item and the conjunction must cover all possible indices of the (bounded) array.
\end{itemize}

These three items are sufficient for generalizing expressions that are part of the array fragment given in Section~\ref{sec:decidability}, which disallows Presburger arithmetic on quantified index-variables, and allows quantified variables to only be used in array read operations. 
\vspace{0.5em}
\begin{definition}
\label{lemma:matching}
Two predicates $\phi_1$ and $\phi_2$ are \emph{matching} predicates if $\phi_1$ contains an array read operation $A[c]$ and $\phi_2$ contains an array read operation $A[d]$, where $c$ and $d$ are constants, and if we replace both $c$ and $d$ with the same fresh variable $z$, $\phi_1$ and $\phi_2$ are equisatisfiable.  
\end{definition}
\vspace{0.5em}
Given a predicate $\phi$ which contains an array read $A[c]$, we use $\phi(z)$ as short-hand for the same predicate with the constant $c$ replaced by a fresh variable. This relationship is transitive, if $\phi_1$ and $\phi_2$ are \emph{matching} predicates, and $\phi_3$ and $\phi_2$ are \emph{matching} predicates then $\phi_1$ and $\phi_3$ are \emph{matching} predicates. It is also commutative. 

A conjunction $C$ over predicates $\phi_0, ..., \phi_n$ can be replaced with a universal quantifier over $\phi(z)$ if the constant array indices we replaced in $\phi_0, ..., \phi_n$ to obtain $\phi(z)$ span the full range of the bounded arrays in the finite-domain specification.
\vspace{0.5em}
\begin{definition}
\label{lemma:conjunctions}
A conjunction $C$ over predicates $\phi_0, ...,\phi_n$ is equisatisfiable with the expression $\forall z.\, \phi_0(z)$, in the finite-domain with bounded arrays, if $\phi_0, ..., \phi_n$ are matching predicates, and the original constants span the full range of the finite-domain bounded arrays.
\end{definition}
\vspace{0.5em}
A similar statement to definition~\ref{lemma:conjunctions} can be written for disjunctions and existential quantifiers. Using definition~\ref{lemma:matching} and definition~\ref{lemma:conjunctions}, we are able to 
apply a procedure to replace conjunctions and disjunctions 
iteratively on the syntax-tree of the finite-domain candidate program $P^b$, starting
at the leaf nodes and working upwards, as shown in Algorithm~\ref{alg:reintroduce_quantifiers}.

\paragraph{Running example: } Consider a possible $P^b$ for our running example, in a finite-domain with arrays of length $2$: $(A[0]\geq 0 \wedge
A[1]\geq 0 \wedge c \geq 0)$. This is a conjunction of predicates:
\begin{align*}
\phi_0 = A[0]>0, \,\, 
\phi_1 = A[1]>0, \,\,
\phi_2 = c>0 
\end{align*}
If we replace the constant indices in the array read operations in $\phi_0$ and $\phi_1$, we can see that the two predicates are \emph{matching}, and the constants span the full range of the finite-domain array. $\phi_2$ does not match any other predicate. We thus replace the first conjunction with a universal quantifier, and the expression becomes $(\forall z\, A[z]>0) \wedge (c>0)$.
% \end{runningexample}

\paragraph{Beyond the decidable array fragment: }
We add two more checks that allow us to handle limited cases outside the decidable array property fragment, specifically we consider the case where Presburger arithmetic is applied to quantified index-variables and where limited cases where quantified index-variables are used outside of array read operations. That is:
\begin{itemize}
  \item if more than one element of the same array is indexed, we look for constant difference relationships between the array elements indexed in the predicate, and check these relationships are the same across all predicates;
  \item and if the predicate contains constants of the same type as the array index that are not used for indexing arrays, we look for constant adjustment relationships between the constants and the array indices, and check if these are the same across all predicates.
\end{itemize}
The formal definitions of these rules can be found in Appendix~\ref{sec:beyond}.

\paragraph{Nested quantifiers: }
Although nested quantifiers are outside of the decidable array fragment, transforming a finite-domain candidate solution $P^b$ to a solution with nested quantifiers 
requires no further transformation rules.
Algorithm~\ref{alg:reintroduce_quantifiers} applies the transformations recursively and, given an expression as input,
 begins by calling itself on all of the operands of that expression, and by doing so is able to introduce nested quantifiers.

\begin{figure}
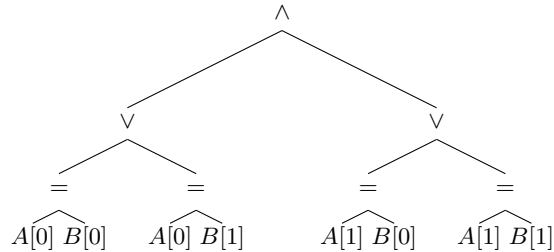

\Tree [.$\wedge$ [.$\vee$ [.= $A[0]$ $B[0]$  ].=  [.= $A[0]$ $B[1]$  ].= !\qsetw{2.25cm} ].$\vee$  [.$\vee$  [.= $A[1]$ $B[0]$ ].= [.= $A[1]$ $B[1]$ ].= !\qsetw{2.25cm} ].$\vee$ !\qsetw{5cm} ].$\wedge$
\caption{Syntax tree example\label{fig:syntaxtree} }
\end{figure}

Consider the expressions $(A[0]=B[0] \vee A[0]=B[1])\wedge (A[1]=B[0] \vee A[1]=B[1])$. 
The syntax tree for this expression is shown in Figure~\ref{fig:syntaxtree}.

%\Tree [.$\wedge$ [.$\vee$ $A[1]=B[0]$ $A[1]=B[1]$ ].$\vee$  [.$\vee$  [.= $A[1]$ $B[0]$ ].= [.= $A[1]$ $B[1]$ ].= ].$\vee$  ].$\wedge$
The key comparison the algorithm makes are:
\begin{enumerate}
  \item Compare the disjunction operands:\\ $(A[0]=B[0])$ and $(A[0]=B[1])$.\\
  %  \begin{itemize}
        %\item 
        Replace with:\\ $\exists z_1 \, A[0]=A[z_1]$.
   %   \end{itemize}
    \item Compare the disjunction operands:\\ $(A[1]=B[0])$ and $(A[1]=B[1])$.\\
   % \begin{itemize}
     % \item 
      Replace with:\\ $\exists z_2 \, A[1]=B[z_2]$.
   % \end{itemize}
    \item Compare the conjunction operands: \\
     $\exists z_1 \, A[1]=B[z_1]$ and $\exists z_2 \, A[0]=B[z_2]$\\
   % \begin{itemize}
     % \item 
      Replace with:  \\  
            $\forall z_3 \exists z_1 \, A[z_3]=B[z_1]$    
   % \end{itemize}

\end{enumerate}

\subsection{Synthesis-based Generalization}
If the syntactic generalization fails, we hypothesize that this is likely because the 
property described by the generalized 
candidate solution does not hold for all array indices. 
We use the predicates found in the generalized
solution as a syntactic template for a synthesis query, which asks whether there exists
a solution to the unbounded specification $\sigma$
that is constructed using these predicates, and additional simple expressions
that can be constructed from non-array parameters, comparison operators, and conjunctions and disjunctions. 
As an exemplar, given a candidate $\forall i, x[i]>0 \vee y=0$, for a program $P$ that accepts three input parameters, an array $x$ and an integer $y$ and $z$, and returns a boolean, we will generate the following syntactic template:
\begin{lstlisting}[style=grammar]
Program::= B
B::= B $\wedge$ B|B $\vee$ B|I $\geq$ I|I $\leq$ I|I = I|(y = 0)| $(\forall i, x[i]>0)$| 
$(\forall i\, ($I$\leq i<$I$)\implies x[i]>0$)
I::= 0|1|y|z|I-I|I+I
\end{lstlisting}
One of the key things this synthesis-based generalization procedure achieves is identifying whether a syntactically generalized solution is a component of a valid solution, if it is, for example, only applied to a subsection of the array that was larger than the original bound. For instance, a valid solution may be $\forall i, (i<z)\implies x[i]>0 \vee y=0$

\paragraph{Extensions: }
The generalization phase is incomplete. There is scope for syntactic generalization of 
further expressions outside of the decidable array fragment. For instance,
Array indices as expressions outside of Presburger arithmetic.

\section{Evaluation}
\label{sec:eval}
\begin{table}
\begin{center}
\begin{tabular}{c| p{3.2cm} p{1.7cm} p{1.7cm} p{1.7cm} p{1.7cm}} 
&Solver            & Z3        & QUIC3   &  CVC4    & \toolname \\\hline\hline
%\multicolumn{5}{c}{SV Comp invariant synthesis problems }    \\\hline
SV-COMP &No. solved           &  1/24     &  7/24   &   1/24   & 10/24   \\
&No. unknown          &  11/24    &  11/24  &   0/24   &  n/a    \\
&No. time-out         &  12/24    &  6/24   &   23/24  &   14    \\ 
&Avg. solving time    &  $<$0.1s  &  $<$0.1s&  $<$0.1s &   122.1s      \\ 
&Median. solving time &  $<$0.1s  &  $<$0.1s&  $<$0.1s &  121.3s        \\ \hline
%\multicolumn{5}{c}{Crafted invariant synthesis problems }    \\ \hline
Crafted Inv&No. solved&  0/19     &  1/19   &  3/19    & 16/19   \\
&No. unknown          &  18/19    &  18/19  &  0/19    &  n/a    \\
&No. time-out         &  1/19     &  0/19   &  16/19   &  3/19   \\
&Avg. solving time    &    n/a    & $<$0.1s &  $<$0.1s &  0.5s    \\
&Median. solving time &  $<$0.1s  &  $<$0.1s&  $<$0.1s &  $<$0.1s  \\ \hline
% Average solving time & $<$0.1s & $<$0.1s & 0.3s& 0.4s\\ \hline
%\multicolumn{5}{c}{Program Sketching} \\\hline
Sketching&No. solved  &  n/a     &  n/a     &  10/22   & 12/22   \\
&No. unknown          &  n/a     &  n/a     &   0/22   &  n/a    \\
&No. time-out         &  n/a     &  n/a     &  12/22   &  10/22  \\ 
&Avg. solving time    &          &          & 1.7s  &   30.5s \\ 
&Median. solving time &  $<$0.1s &  $<$0.1s &  $<$0.1s &  0.5s   \\ \hline\hline
Total&Total solved         &   1      &  9  &   14     &   38    \\ 
&Avg. solving time    & $<$0.1s  & $<$0.1s  & $<$0.1s  &   40.8s   \\ 
&Median. solving time &  $<$0.1s &  $<$0.1s &  $<$0.1s &  0.1s   \\ \hline\hline
\end{tabular}
\end{center}
\caption{Examples solved by each solver. We ran the experiments with a 300\,s 
timeout. We differentiate between unsolved benchmarks that time-out and unsolved 
benchmarks where the solver returns unknown. \toolname does not implement any way of returning unknown.}
\end{table}
% \todo[inline]{Update results and add paragraph on JavaStringUtils benchmarks}
We implement \toolname using CVC4 version 1.9[pre-release] as the synthesis phase. 
We use Z3 version 4.8.7 for verification. 
The communication between the transformation phases and the synthesis phase is done in SyGuS-IF, allowing any existing SyGuS solver to be substituted into the synthesis phase. Furthermore, the verification phase produces standard SMT-lib, allowing any existing SMT solver to be used as a back-end. 

We evaluate our algorithm on $65$ benchmarks: $24$ invariant synthesis benchmarks adapted from the 
Software Verification Competition~\cite{10.1007/978-3-030-45237-7_21}; $19$ example challenging invariant synthesis problems, crafted to test the capabilities of our algorithm, and $22$ program sketching problems, $19$ of which are adapted from the Java StringUtils Class. The Java StringUtils Class is a good target for program sketching problems since the functions reason
about strings, which are arrays of characters. We represent these as arrays of integers. Furthermore, each function is provided with a
specification in the source code. All benchmarks and code are available to download\footnote{\url{https://drive.google.com/file/d/1Y__q0CPTDLZ5swQZfnUw7R-brjOhTz1_}}.
 All but $7$ of our benchmarks use additional quantifiers beyond the standard program synthesis formulation. The $7$ benchmarks that do not are taken from the program sketching JavaStringUtils benchmarks.

We run QUIC3~\cite{DBLP:conf/atva/GurfinkelSV18} and the
Z3~\cite{DBLP:journals/fmsd/KomuravelliGC16} Horn Solver, both contained
within Z3 version 4.8.7, on the examples that can be expressed in 
CHC-format~\cite{DBLP:journals/corr/abs-2008-02939}, and 
CVC4~\cite{DBLP:journals/corr/abs-1806-08775} version 1.9[pre-release] on all benchmarks
with a time-out of 300s.
None of these solvers officially support this combination of quantification, and so
are only able to solve a subset of these benchmarks.
% We note that neither
% FreqHorn~\cite{DBLP:conf/cav/FedyukovichPMG19}, nor
% CVC4~\cite{DBLP:conf/cav/ReynoldsBNBT19} officially support the
% quantification in these specifications and were unable to solve any of these
% benchmarks. 
\toolname solves more than double the number of examples than any other tool. The median solving time
is comparable to the other solvers, but the average is larger since some benchmarks that go through several iterations
to find a model size large enough may have to wait for the synthesis phases to time-out on the earlier iterations.
The synthesis time-out is a heuristic and with a greater time-out we may be able to solve more benchmarks
but slower.
It typically takes between $1$ and $4$ iterations to find a model size large enough to solve each benchmark, and $23$ of the benchmarks were solved by the syntactic generalization procedure, and a further $15$ required the synthesis based generalization. 
We solve $27$ invariant synthesis benchmarks; 
$18$ of the solutions have single quantifiers and $5$ have alternating quantifiers.
QUIC3 is able to solve $8$ invariant synthesis benchmarks, which all have single quantifiers.

CVC4 is able to solve $10$ of the program sketching benchmarks, $2$ of which require quantifiers in the specification but not in the synthesized solutions, and typically where the benchmark is single invocation (that is, where the function is called only with the same arguments in the same order), for which CVC4 contains special purpose algorithms~\cite{DBLP:conf/cav/ReynoldsDKTB15}. \toolname is able to solve $4$ benchmarks that CVC4 is unable to solve due to the size of the array that must be reasoned about, but misses solving one benchmark that has a single invocation solution. A number of the sketching benchmarks would be solvable by \toolname if we included helper functions from the benchmark in the syntactic template for the bounded synthesis problem.
 
% \paragraph{Linear relationships between array indices: } 
% Firstly, we were unable to solve some examples due to constant difference relationships existing between the array indices that demanded a larger $\sigma^b$ than we could solve. For example, 
% an invariant which requires an array element at index $i$ to be equal to an array element at index $i+10$ requires a finite-domain specification that allows arrays at least $11$ elements long. We could address this challenge by refining the restriction phase of our algorithm.
\subsection{Threats to validity} 

\vspace{1em}

\textit{Benchmark selection: }We report an assessment of our approach over a set of real-world(SV-COMP and Java StringUtils sketching) and crafted benchmarks designed to test the capability of our algorithm, since the synthesis community is currently lacking a standard set of benchmarks containing arrays and quantifiers. 

\noindent \textit{Dependency on CVC4: }Our algorithm depends on the abilities of CVC4, or another synthesis solver, to solve the specification $\sigma^b$. For some benchmarks, where the verification query would fall outside the decidable fragment identified, CVC4 was unable to solve the smallest model we were able to generate within the timeout. Since the actual solutions for these small models are short (typically $3-4$ operations, reasoning about $2-4$ array elements), we believe that a valuable direction for future work would be exploring enumerative
techniques tailored to these types of problems where the search space for an enumerative engine is small. 
These are shown as unsolved by \toolname in the results table. However, in order to validate our generalization procedure, we also ran experiments where mocked the expected result from CVC4 and showed that our generalization process is capable of producing the correct result.

\noindent\textit{Modeling of loops in benchmarks:} Nested or multiple loops are represented in some of our benchmarks using quantifiers. This is an 
easy representation for a human modeling such benchmarks to understand. 
However, one could equally express these benchmarks using multiple loops, 
reducing the number of quantifiers needed, but requiring the an invariant synthesis
tool to synthesize several inter-dependent invariants. We experimented with this
representation and found that the results of QUIC3 and Z3 were not affected
by our choice of representation.

\section{Related work}
\label{sec:related}
Program sketching was originally presented by Solar-Lezama et al.~\cite{DBLP:conf/asplos/Solar-LezamaTBSS06}. 
The Sketch tool allows synthesis of expressions, if the user encodes a grammar
as part of a generator function, and it performs well when the user is able 
to provide a sufficiently detailed sketch, but it does not support quantifiers and only has 
support for limited loops. Many of the SyGuS competition benchmarks in the General
Track are equivalent to program sketching, albeit without arrays, and the leading solver in this track
for 2018 and 2019 was CVC4~\cite{DBLP:journals/corr/abs-1806-08775}, which 
we compare to in our experimental evaluation.

The community has invested a large amount of effort into the problem
of invariant synthesis, which is a specific instance of the program synthesis
problem we tackle,
and identified a broad variety of special cases in which reachability
properties of parametric systems are decidable. For the unrestricted
case, the community has devised numerous heuristic methods for guessing
and possibly refining the 
predicate~$I$~\cite{DBLP:conf/aplas/KongJDWY10, DBLP:conf/kbse/NguyenDV17,
DBLP:conf/nips/SiDRNS18}. 
CVC4 and LoopInvGen~\cite{pldi/2016/PadhiSM} both perform well in the syntax-guided
synthesis competition, but neither can handle quantifiers in synthesis.

There are many approaches that synthesize invariants containing quantifiers
over array indices, however, none of them allow for quantification in the
specification.  QUIC3~\cite{DBLP:conf/atva/GurfinkelSV18} is an adaptation of
IC3~\cite{DBLP:conf/vmcai/Bradley11} to synthesize quantified invariants.  
Larraz et al.~\cite{DBLP:conf/vmcai/LarrazRR13}
present an SMT-based array invariant generation approach, which is limited
to universally quantified loop invariants over arrays and scalar variables.
FreqHorn~\cite{DBLP:conf/cav/FedyukovichPMG19} uses syntax-guided synthesis
to synthesize quantified invariants: they identify potentially useful facts 
about elements accessed
in a loop and attempt to generalize these to hypothesis about the
entire range of the variables.  This is the approach most similar to our
work, however the way they identify the range of elements is specific to a
loop invariant synthesis problem.  Our approach relies on a more general
program synthesis phase to identify useful elements and so is not restricted
to loop invariant synthesis.  FreqHorn also does not permit additional quantification in
the specification and we so are unable to compare to this tool. 
There also exists recent work on using trace-logic
to verify loops with properties that use alternating 
quantifiers~\cite{DBLP:journals/corr/abs-2008-01387}; the approach
is specific to loop verification, and 
cannot tackle the general program synthesis problem,
although it can verify some loops that have properties containing alternating quantifiers. 
We do not compare directly to the tool since they cannot tackle the general synthesis problem,
and do not accept SyGuS-IF or CHC format, but we would expect them to 
perform comparably to \toolname on the invariant synthesis benchmarks
but be unable to tackle the general program synthesis problems.

I4~\cite{DBLP:conf/sosp/MaGJKKS19} is an algorithm that uses a similar insight
based on finding invariants for small instances of protocols using model-checking,
 and generalizing
them to larger numbers of nodes. Since the approach is based on model-checking, it 
is limited to invariant generation, whereas our approach can handle more
general synthesis cases. They also handle only universal 
quantifiers over nodes of the distributed protocol, and not quantifier 
alternations or existential quantifiers.

% Older work exists which cannot handle quantification and generates invariants based on
% templates~\cite{DBLP:conf/aplas/KongJDWY10} and using predicate
% abstraction~\cite{DBLP:conf/vmcai/LahiriB04}. Ours, in contrast, uses a syntax-guided synthesis approach.

Our algorithm is in part inspired by verification approaches which use the
principle of abstracting a verification problem by considering short
versions of bit-vectors and arrays~\cite{DBLP:conf/fmcad/SinhaSMSW12,DBLP:journals/sttt/BryantKOSSB09}.
Khasidashvili et al.~\cite{DBLP:conf/fmcad/KhasidashviliKV09} verify equivalences of memory
by translation into first-order logic, and note that for some specific designs
this falls into a decidable fragment.
Verification procedures such as CEGAR~\cite{DBLP:journals/jacm/ClarkeGJLV03} iteratively refine
an abstraction, and we iteratively refine $\sigma^b$. A key difference is that CEGAR relies on
refining the abstraction until it the abstraction is precise enough that a counterexample is valid on the original program. We
only refine $\sigma^b$ until it is precise enough that a satisfying assignment $P^b$ can be generalized
to be a valid solution $P$ for the original specification. 
The restricted specification $\sigma^b$ is almost never precise enough that
$P^b$ is a valid solution for $\sigma$.

% Some of our motivating examples are based on synthesizing invariants for arrays. 
% However, there are methods for verifying array programs without using loop
% invariants: Abstraction of the array to a fixed number of elements is used
% to reduce array modifying loops with unknown bounds to loops with a known,
% small
% bound~\cite{DBLP:conf/lopstr/JanaKDVC16,DBLP:conf/tacas/KumarSVS18,DBLP:conf/sas/MonniauxA15}. 
% An imprecise approach involves
% abstracting the array so that all array elements appear in a single memory
% location~\cite{DBLP:journals/ftpl/BertraneCCFMMR15}.  Under-approximating
% loops in programs by
% acceleration~\cite{DBLP:conf/cav/BozgaHIKV09,DBLP:journals/fmsd/KroeningLW15}
% may also remove the need for invariants but since it approximates the loops
% the result is not guaranteed to be correct. 

\section{Conclusions}
We have presented an algorithm which can synthesize expressions containing 
alternating quantifiers, to specifications containing quantification over arrays. 
The synthesis algorithm works by bounding unrestricted domains in the synthesis specification, 
synthesizing a solution to this finite-domain specification, and then attempting to generalize the 
solution to that to the unrestricted domain. 
We are able to synthesize quantified expressions that elude 
existing solvers and, despite being a general program synthesis algorithm, perform well against
specialized invariant synthesis solvers.
Furthermore, our algorithm is a framework that exploits the strengths of existing 
state-of-the-art solvers,
and so as the speed and scalability of quantifier-free syntax-guided synthesis improves, so will the 
performance of our algorithmic framework. 

\bibliography{arrays}{}
\bibliographystyle{plain}

%\appendix
%\input{benchmarks.tex}
\appendix
\section{Syntactic generalization beyond the array fragment}
\label{sec:beyond}
Recall, we wish to generalize limited cases outside the decidable array property fragment. Specifically we consider the case where Presburger arithmetic is applied to quantified index-variables and where limited cases where quantified index-variables are used outside of array read operations. That is:
\begin{itemize}
  \item if more than one element of the same array is indexed, we look for constant difference relationships between the array elements indexed in the predicate, and check these relationships are the same across all predicates;
  \item and if the predicate contains constants of the same type as the array index that are not used for indexing arrays, we look for constant adjustment relationships between the constants and the array indices, and check if these are the same across all predicates.
\end{itemize}

We extend our definition of \emph{matching} predicates to allow constant difference relationships between the array elements indexed in the predicate:
\vspace{0.5em}
\begin{definition}
\label{lemma:matching_extended}
Two predicates $\phi_1$ and $\phi_2$ are \emph{matching} predicates if $\phi_1$ contains array read operations 
$A[c_0],..., A[c_n]$ and $\phi_2$ contains array read operations $A[d_0],..., A[d_n]$, and if we replace $c_0, ..., c_n$
and $d_0, ..., d_n$ with the same set of expressions $z+e_0, ..., z+e_n$ where $z$ is the same fresh variable 
and $e_0,...,e_n$ is a set of constants, and $\phi_1$ and $\phi_2$ are equisatisfiable.  
\end{definition}
\vspace{0.5em}
We use $\phi(z)$ to indicate the expression obtained by replacing multiple array read operations in $\phi$ with a set of expressions $z+e_0, ..., z+e_n$. We create a similar rule for constants of the same type as the array indices, that are used outside of array read operations. 
\vspace{0.5em}
\begin{definition}
\label{lemma:matching_constants}
Two predicates $\phi_1$ and $\phi_2$ contain array read operations 
$A[c_0],..., A[c_n]$ and constants $x_0, ..., x_n$,
and $\phi_2$ contains array read operations $A[d_0],..., A[d_n]$ and constants $y_0, ..., y_n$. 
We replace $c_0,...,c_n$
and $d_0, ..., d_n$ with the same set of expressions $z+e_0, ..., z+e_n$ where $z$ is the same fresh variable and $e_0, .., e_n$ is a set of constants. We replace $x_0,...,x_n$ and $y_0,...,y_n$ with $z+f_0,...,z+f_n$ where $z$ is the same variable as before, and $f_0, ..., f_n$ is a set of constants.
If $\phi_1$ and $\phi_2$ are now equisatisfiable, the two predicates are \emph{matching} predicates.  
\end{definition}
\vspace{0.5em}
We use $\phi(z)$ to indicate the expression obtained by replacing multiple array read operations in $\phi$ with a set of expressions $z+e_0, ..., z+e_n$ and multiple constants with$z+f_0,...,z+f_n$. A conjunction $C$ that reasons about predicates $\phi_0, ..., \phi_n$ can be replaced with the expression $\forall z ., \phi_0(z)$ if $\phi_0, ..., \phi_n$ are \emph{matching} predicates and the constants that we replaced with $z+0$ span the full range of the restricted domain. That is, definition~\ref{lemma:conjunctions} still applies. 
Consider the example $(A[0]<A[1])\wedge(A[1]<A[2])$, in the finite-domain of arrays of length 2:
\begin{align*}
\phi_0 = (A[0]<A[1]), \,\, 
\phi_1 = (A[1]<A[2]), \,\,
\end{align*}
\noindent If we replace the array reads with $A[z+0], A[z+1]$, then, by definition~\ref{lemma:matching_extended} the two predicates are \emph{matching}, and the constants that we replaced by $z+0$ span the full range of the restricted domain. We can thus replace the conjunction with the expression $\forall z'\,A[z]< A[z+1]$.

\end{document}

%% file: commands.tex
\definecolor{codegreen}{cmyk}{81,0,39,0}
\definecolor{codegray}{cmyk}{45,25,16,59}
\definecolor{codepurple}{cmyk}{76,34,21,0}
\definecolor{backcolour}{rgb}{1,1,1}
\definecolor{codemagenta}{cmyk}{0,85,88,0}

\lstdefinelanguage{uclid}{
  sensitive = true,
  keywords={module, forall, exists, Lambda, if, else, assert, assume, havoc,
            for, range, skip, case, esac, init, next, control, function, procedure,
            returns, call, define, type, var, input, output, const, property,
            invariant, synthesis, grammar, requires, ensures, modifies, instance, axiom, 
            enum, record, integer, boolean, true, false},
  numbers=left,
  numberstyle=\footnotesize,
  stepnumber=1,
  numbersep=9pt,
  showstringspaces=false,
  breaklines=true,
  comment=[l]{//},
  morecomment=[s]{/*}{*/},
}

\lstdefinestyle{uclidstyle}{
  commentstyle=\color{codegreen},
  keywordstyle=\color{codemagenta},
  numberstyle=\tiny\color{codegray},
  stringstyle=\color{codepurple},
  breakatwhitespace=false,
  basicstyle=\scriptsize\ttfamily,
  breaklines=true,
  captionpos=b,
  keepspaces=true,
  showspaces=false,
  showtabs=false,
  tabsize=2,
  frame=single,
  aboveskip=0em,
  belowskip=0em,
}

%% file: new_alg.tikz
\begin{tikzpicture}[>=latex,x=3cm,y=2cm]

\node[rectangle,draw,rounded corners=2ex,minimum height=1.2cm,minimum width=2.5cm,align=center] at (-0.5,0) (t){Restrict\\ $\sigma \rightarrow \sigma^b$} ;

\node[rectangle,draw,dashed,rounded corners=2ex,minimum height=1.2cm,minimum width=2.5cm,align=center] at (0.5,-1) (b){increase\\ $b$} ;

\node[rectangle,draw,rounded corners=2ex,minimum height=1.2cm,minimum width=2.5cm,align=center] at (0.5,0) (synth){Synthesize \\ $\exists P^b \forall x\, \sigma^b(P^b, x)$};

\node[rectangle,draw,rounded corners=2ex,minimum height=1.2cm,minimum width=2.5cm,align=center] at (1.5,0) (r1){Generalize \\ (syntactic)\\$P^b \rightarrow P^*$};

\node[rectangle,draw,rounded corners=2ex,minimum height=1.2cm,minimum width=2.5cm,align=center] at (1.5,-1) (v){Verify\\\ $\exists x \,\neg \sigma(P^*,x)$};

\node[rectangle,draw,rounded corners=2ex,minimum height=1.2cm,minimum width=2.5cm,align=center] at (2.5,-1) (r2){Generalize \\ (synthesis-based)\\$P^b \rightarrow P^*$};
\node [align=center] at (2.5,0) (soln) {\textbf{found solution}\\$P$};
\node [align=center] at (-1.2,0)(start){};
\node at(-0.8, -0.1)(corner){};

\path[->] (t) edge[above] node [yshift=0.1cm]{$\sigma^b$} (synth);

\path[->] (synth) edge[above] node {$P^b$} (r1);
\path[->] (r1) edge[left] node []{$P^*$} (v);
\path[->] (v) edge[above] node []{$P^*$} (r2);

\path[->] (r2) edge[above] node [xshift=0.1cm]{} (soln);

\path[->] (v) edge[above] node [align=center]{} (soln);

%\path[->] (v) edge[below] node [xshift=-0.5cm]{} (b);
\path[->] (r2) edge[below, bend left] node []{no solution found} (b);
\path[->] (b) edge[below, bend left] node [xshift=-0.5cm]{} (t);

\path[->] (synth) edge [left] node [align=center]{no soln\\found} (b);

\path[->] (start) edge[above] node {}(t);

\end{tikzpicture}

%% file: paper.bbl
\begin{thebibliography}{10}

\bibitem{DBLP:journals/corr/abs-1711-11438}
Rajeev Alur, Dana Fisman, Rishabh Singh, and Armando Solar{-}Lezama.
\newblock {SyGuS-Comp} 2017: Results and analysis.
\newblock {\em CoRR}, abs/1711.11438, 2017.

\bibitem{sygus}
Rajeev Alur, Dana Fisman, Rishabh Singh, and Abhishek Udupa.
\newblock Syntax guided synthesis competition.
\newblock \url{http://sygus.seas.upenn.edu/SyGuS-COMP2017.html}, 2017.

\bibitem{DBLP:journals/corr/abs-1806-08775}
Clark~W. Barrett, Haniel Barbosa, Martin Brain, Duligur Ibeling, Tim King, Paul
  Meng, Aina Niemetz, Andres N{\"{o}}tzli, Mathias Preiner, Andrew Reynolds,
  and Cesare Tinelli.
\newblock {CVC4} at the {SMT} competition 2018.
\newblock {\em CoRR}, abs/1806.08775, 2018.

\bibitem{DBLP:journals/ftpl/BertraneCCFMMR15}
Julien Bertrane, Patrick Cousot, Radhia Cousot, J{\'{e}}r{\^{o}}me Feret,
  Laurent Mauborgne, Antoine Min{\'{e}}, and Xavier Rival.
\newblock Static analysis and verification of aerospace software by abstract
  interpretation.
\newblock {\em Foundations and Trends in Programming Languages},
  2(2-3):71--190, 2015.

\bibitem{10.1007/978-3-030-45237-7_21}
Dirk Beyer.
\newblock Advances in automatic software verification: {SV-COMP} 2020.
\newblock In Armin Biere and David Parker, editors, {\em Tools and Algorithms
  for the Construction and Analysis of Systems}, pages 347--367, Cham, 2020.
  Springer International Publishing.

\bibitem{DBLP:conf/cav/BozgaHIKV09}
Marius Bozga, Peter Habermehl, Radu Iosif, Filip Konecn{\'{y}}, and Tom{\'{a}}s
  Vojnar.
\newblock Automatic verification of integer array programs.
\newblock In {\em {CAV}}, volume 5643 of {\em Lecture Notes in Computer
  Science}, pages 157--172. Springer, 2009.

\bibitem{DBLP:conf/vmcai/Bradley11}
Aaron~R. Bradley.
\newblock Sat-based model checking without unrolling.
\newblock In {\em {VMCAI}}, volume 6538 of {\em Lecture Notes in Computer
  Science}, pages 70--87. Springer, 2011.

\bibitem{DBLP:conf/vmcai/BradleyMS06}
Aaron~R. Bradley, Zohar Manna, and Henny~B. Sipma.
\newblock What's decidable about arrays?
\newblock In {\em {VMCAI}}, volume 3855 of {\em Lecture Notes in Computer
  Science}, pages 427--442. Springer, 2006.

\bibitem{DBLP:journals/sttt/BryantKOSSB09}
Randal~E. Bryant, Daniel Kroening, Jo{\"{e}}l Ouaknine, Sanjit~A. Seshia, Ofer
  Strichman, and Bryan~A. Brady.
\newblock An abstraction-based decision procedure for bit-vector arithmetic.
\newblock {\em {STTT}}, 11(2):95--104, 2009.

\bibitem{DBLP:journals/corr/CaulfieldRST15}
Benjamin Caulfield, Markus~N. Rabe, Sanjit~A. Seshia, and Stavros Tripakis.
\newblock What's decidable about syntax-guided synthesis?
\newblock {\em CoRR}, abs/1510.08393, 2015.

\bibitem{DBLP:journals/jacm/ClarkeGJLV03}
Edmund~M. Clarke, Orna Grumberg, Somesh Jha, Yuan Lu, and Helmut Veith.
\newblock Counterexample-guided abstraction refinement for symbolic model
  checking.
\newblock {\em J. {ACM}}, 50(5):752--794, 2003.

\bibitem{DBLP:conf/tacas/MouraB08}
Leonardo~Mendon{\c{c}}a de~Moura and Nikolaj Bj{\o}rner.
\newblock {Z3:} an efficient {SMT} solver.
\newblock In {\em {TACAS}}, volume 4963 of {\em Lecture Notes in Computer
  Science}, pages 337--340. Springer, 2008.

\bibitem{chc}
Grigory Fedyukovich.
\newblock {CHC} competition.
\newblock \url{https://chc-comp.github.io/2018/format.html}, 2018.

\bibitem{DBLP:conf/cav/FedyukovichPMG19}
Grigory Fedyukovich, Sumanth Prabhu, Kumar Madhukar, and Aarti Gupta.
\newblock Quantified invariants via syntax-guided synthesis.
\newblock In {\em {CAV} {(1)}}, volume 11561 of {\em Lecture Notes in Computer
  Science}, pages 259--277. Springer, 2019.

\bibitem{DBLP:journals/corr/abs-2008-01387}
Pamina Georgiou, Bernhard Gleiss, and Laura Kov{\'{a}}cs.
\newblock Trace logic for inductive loop reasoning.
\newblock {\em CoRR}, abs/2008.01387, 2020.

\bibitem{DBLP:conf/atva/GurfinkelSV18}
Arie Gurfinkel, Sharon Shoham, and Yakir Vizel.
\newblock Quantifiers on demand.
\newblock In {\em {ATVA}}, volume 11138 of {\em Lecture Notes in Computer
  Science}, pages 248--266. Springer, 2018.

\bibitem{DBLP:conf/lopstr/JanaKDVC16}
Anushri Jana, Uday~P. Khedker, Advaita Datar, R.~Venkatesh, and Niyas C.
\newblock Scaling bounded model checking by transforming programs with arrays.
\newblock In {\em {LOPSTR}}, volume 10184 of {\em Lecture Notes in Computer
  Science}, pages 275--292. Springer, 2016.

\bibitem{DBLP:conf/fmcad/KhasidashviliKV09}
Zurab Khasidashvili, Mahmoud Kinanah, and Andrei Voronkov.
\newblock Verifying equivalence of memories using a first order logic theorem
  prover.
\newblock In {\em {FMCAD}}, pages 128--135. {IEEE}, 2009.

\bibitem{DBLP:journals/fmsd/KomuravelliGC16}
Anvesh Komuravelli, Arie Gurfinkel, and Sagar Chaki.
\newblock {SMT}-based model checking for recursive programs.
\newblock {\em Formal Methods in System Design}, 48(3):175--205, 2016.

\bibitem{DBLP:conf/aplas/KongJDWY10}
Soonho Kong, Yungbum Jung, Cristina David, Bow{-}Yaw Wang, and Kwangkeun Yi.
\newblock Automatically inferring quantified loop invariants by algorithmic
  learning from simple templates.
\newblock In {\em {APLAS}}, volume 6461 of {\em Lecture Notes in Computer
  Science}, pages 328--343. Springer, 2010.

\bibitem{DBLP:journals/fmsd/KroeningLW15}
Daniel Kroening, Matt Lewis, and Georg Weissenbacher.
\newblock Under-approximating loops in {C} programs for fast counterexample
  detection.
\newblock {\em Formal Methods in System Design}, 47(1):75--92, 2015.

\bibitem{DBLP:series/txtcs/KroeningS16}
Daniel Kroening and Ofer Strichman.
\newblock {\em Decision Procedures - An Algorithmic Point of View, Second
  Edition}.
\newblock Texts in Theoretical Computer Science. An {EATCS} Series. Springer,
  2016.

\bibitem{DBLP:conf/tacas/KumarSVS18}
Shrawan Kumar, Amitabha Sanyal, R.~Venkatesh, and Punit Shah.
\newblock Property checking array programs using loop shrinking.
\newblock In {\em {TACAS} {(1)}}, volume 10805 of {\em Lecture Notes in
  Computer Science}, pages 213--231. Springer, 2018.

\bibitem{DBLP:conf/vmcai/LarrazRR13}
Daniel Larraz, Enric Rodr{\'{\i}}guez{-}Carbonell, and Albert Rubio.
\newblock {SMT}-based array invariant generation.
\newblock In {\em {VMCAI}}, volume 7737 of {\em Lecture Notes in Computer
  Science}, pages 169--188. Springer, 2013.

\bibitem{DBLP:conf/sosp/MaGJKKS19}
Haojun Ma, Aman Goel, Jean{-}Baptiste Jeannin, Manos Kapritsos, Baris Kasikci,
  and Karem~A. Sakallah.
\newblock {I4:} incremental inference of inductive invariants for verification
  of distributed protocols.
\newblock In {\em {SOSP}}, pages 370--384. {ACM}, 2019.

\bibitem{DBLP:conf/sas/MonniauxA15}
David Monniaux and Francesco Alberti.
\newblock A simple abstraction of arrays and maps by program translation.
\newblock In {\em {SAS}}, volume 9291 of {\em Lecture Notes in Computer
  Science}, pages 217--234. Springer, 2015.

\bibitem{DBLP:conf/kbse/NguyenDV17}
ThanhVu Nguyen, Matthew~B. Dwyer, and Willem Visser.
\newblock {SymInfer}: inferring program invariants using symbolic states.
\newblock In {\em {ASE}}, pages 804--814. {IEEE} Computer Society, 2017.

\bibitem{pldi/2016/PadhiSM}
Saswat Padhi, Rahul Sharma, and Todd~D. Millstein.
\newblock Data-driven precondition inference with learned features.
\newblock In {\em Proceedings of the 37th {ACM} {SIGPLAN} Conference on
  Programming Language Design and Implementation, {PLDI} 2016, Santa Barbara,
  CA, USA, June 13-17, 2016}, pages 42--56, 2016.

\bibitem{DBLP:conf/cav/ReynoldsBNBT19}
Andrew Reynolds, Haniel Barbosa, Andres N{\"{o}}tzli, Clark~W. Barrett, and
  Cesare Tinelli.
\newblock cvc4sy: Smart and fast term enumeration for syntax-guided synthesis.
\newblock In {\em {CAV} {(2)}}, volume 11562 of {\em Lecture Notes in Computer
  Science}, pages 74--83. Springer, 2019.

\bibitem{DBLP:conf/cav/ReynoldsDKTB15}
Andrew Reynolds, Morgan Deters, Viktor Kuncak, Cesare Tinelli, and Clark~W.
  Barrett.
\newblock Counterexample-guided quantifier instantiation for synthesis in
  {SMT}.
\newblock In {\em {CAV} {(2)}}, volume 9207 of {\em Lecture Notes in Computer
  Science}, pages 198--216. Springer, 2015.

\bibitem{DBLP:journals/corr/abs-2008-02939}
Philipp R{\"{u}}mmer.
\newblock Competition report: {CHC-COMP-20}.
\newblock In Laurent Fribourg and Matthias Heizmann, editors, {\em Proceedings
  8th International Workshop on Verification and Program Transformation and 7th
  Workshop on Horn Clauses for Verification and Synthesis, VPT/HCVS@ETAPS 2020
  2020, and 7th Workshop on Horn Clauses for Verification and SynthesisDublin,
  Ireland, 25-26th April 2020}, volume 320 of {\em {EPTCS}}, pages 197--219,
  2020.

\bibitem{DBLP:conf/nips/SiDRNS18}
Xujie Si, Hanjun Dai, Mukund Raghothaman, Mayur Naik, and Le~Song.
\newblock Learning loop invariants for program verification.
\newblock In {\em NeurIPS}, pages 7762--7773, 2018.

\bibitem{DBLP:conf/fmcad/SinhaSMSW12}
Rohit Sinha, Cynthia Sturton, Petros Maniatis, Sanjit~A. Seshia, and David~A.
  Wagner.
\newblock Verification with small and short worlds.
\newblock In {\em {FMCAD}}, pages 68--77. {IEEE}, 2012.

\bibitem{DBLP:conf/asplos/Solar-LezamaTBSS06}
Armando Solar{-}Lezama, Liviu Tancau, Rastislav Bod{\'{\i}}k, Sanjit~A. Seshia,
  and Vijay~A. Saraswat.
\newblock Combinatorial sketching for finite programs.
\newblock In {\em {ASPLOS}}, pages 404--415. {ACM}, 2006.

\end{thebibliography}
